\documentclass[lettersize,journal]{IEEEtran}
\usepackage{amsmath,amsfonts}
\usepackage{algorithmic}
\usepackage{algorithm}
\usepackage{array}
\usepackage[caption=false,font=normalsize,labelfont=sf,textfont=sf]{subfig}
\usepackage{textcomp}
\usepackage{stfloats}
\usepackage{url}
\usepackage{verbatim}
\usepackage{graphicx}
\usepackage{xcolor}
\usepackage{cite}
\begin{document}

\title{Image Watermarking of Generative Diffusion Models}

\author{Yunzhuo Chen\textsuperscript{1},Jordan Vice\textsuperscript{1}, Naveed Akhtar\textsuperscript{2,1}, Nur Al Hasan Haldar\textsuperscript{3,1}, Ajmal Mian\textsuperscript{1}\\
\small \textsuperscript{1}The University of Western Australia, Perth, Australia \\
\small \textsuperscript{2}The University of Melbourne, Melbourne, Australia \\
\small \textsuperscript{3}Curtin University, Perth, Australia \\
\texttt{\small yunzhuo.chen@research.uwa.edu.au},
\texttt{\small naveed.akhtar1@unimelb.edu.au}, \\
\texttt{\small  nur.haldar@curtin.edu.au },
\texttt{\small jordan.vice@uwa.edu.au},
\texttt{\small ajmal.mian@uwa.edu.au}
\thanks{This research was supported by National Intelligence and Security Discovery Research Grants (project$\#$ NS220100007), funded by the Department of Defence Australia. Professor Ajmal Mian is the recipient of an Australian Research Council Future Fellowship Award (project number FT210100268) funded by the Australian Government.
Dr.~Naveed Akhtar is the recipient of an Australian Research Council Discovery Early Career Researcher Award (project number DE230101058) funded by the Australian Government.}
\thanks{Manuscript received ; revised }}

\markboth{Journal of \LaTeX\ Class Files,~Vol.~14, No.~8, August~2021}%
{Shell \MakeLowercase{\textit{et al.}}: A Sample Article Using IEEEtran.cls for IEEE Journals}

\vspace{10mm}
\IEEEoverridecommandlockouts
\vspace{0.2cm}
\IEEEpubid{0000--0000/00\$00.00~\copyright~2021 IEEE}

\maketitle

\begin{abstract}

Embedding watermarks into the output of generative models is essential for establishing copyright and verifiable ownership over the generated content. Emerging diffusion model watermarking methods either embed watermarks in the frequency domain 
or offer limited versatility of the watermark patterns in the image space, which allows simplistic detection and removal of the watermarks from the generated content.    
To address this issue, we propose a watermarking technique that embeds watermark features into the diffusion model itself. Our technique enables training of a paired watermark extractor for a generative model that is learned through an end-to-end process. The extractor forces the generator, during training,  to 
effectively embed versatile, imperceptible watermarks in the generated content 
while simultaneously ensuring 
their precise recovery. We demonstrate highly accurate watermark embedding/detection and show that it is also possible to distinguish between different watermarks embedded with our method to differentiate between generative models. 



%

\end{abstract}
\begin{IEEEkeywords}

Diffusion model, AI generation, Watermarking, Copyright 

\end{IEEEkeywords}

\section{Introduction}

\label{intro1}

\IEEEPARstart{D}{iffusion} models have recently provided cutting-edge advancement in the quality of generated visual content~\cite{cao2022survey}. They are claimed to have overcome challenges such as, matching posterior distributions in VAEs~\cite{doersch2016tutorial}, managing unpredictability in GANs~\cite{creswell2018generative}, the high computational demands of Markov Chain Monte Carlo techniques in EBMs \cite{geyer1992practical}, and network limitations of normalized flow methods. However, the high quality of diffusion-generated content raises many ethical concerns, including copyright issues~\cite{thankappan2024distributed,elberri2024cyber}. Watermarking the generated image can allow users to trace the image's source and solve issues pertaining to copyright abuse. 
Several watermarking methods have been proposed to protect the copyright of generative models~\cite{binkowski2018demystifying,heusel2017gans,salimans2016improved,zhou2019hype}. However, these methods have their drawbacks. For instance, some approaches \cite{wen2023tree,cox2007digital}  may fail to provide robustness against advanced image transformations such as geometric distortions, sophisticated noise addition, or adversarial attacks~\cite{akhtar2018threat}. Furthermore, the hidden information in watermarked images remains sensitive to simple image transformations such as compression, cropping, and noise addition. This sensitivity can lead to the loss or corruption of the watermark, thereby reducing the effectiveness of the method in maintaining the integrity and traceability of the generated images. 

\begin{figure}
\centering
  \includegraphics[width=0.5\textwidth]{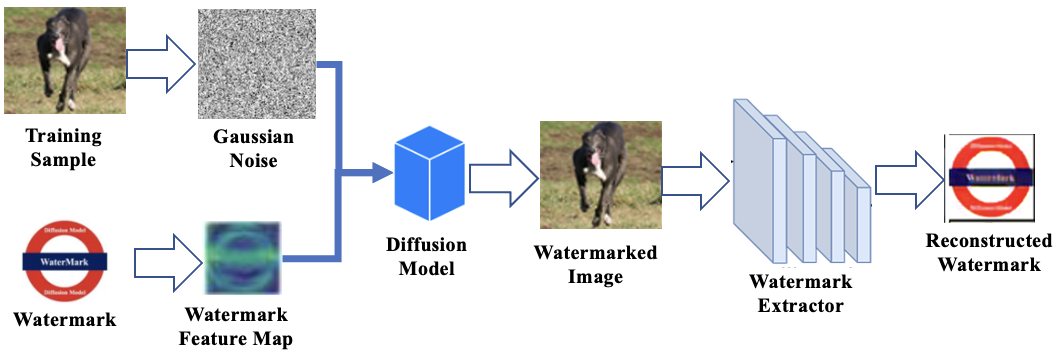}
\caption{\textcolor{black}{During our model training phase, feature maps extracted from a watermark image are combined with a Gaussian sample to train the diffusion model. Generated images obtained from reverse diffusion process contain watermark features. The Watermark Extractor extracts the watermark features from the image to reconstruct the watermark.}} 
\label{fig:1}       
\end{figure}

Three key attributes are often considered when evaluating watermarking techniques:  robustness, imperceptibility and capacity \cite{zhu2018hidden}. Robustness characterizes the preservation of hidden information against image processing attacks. Imperceptibility defines a technique's ability to make the watermarked image indistinguishable from unwatermarked images. Capacity describes the amount of hidden information in a watermark. The watermark may include details such as the copyright holder's name or other identifying information. 

A \textit{blind} watermarking method embeds hidden information within an image, making it invisible to the naked eye and detectable without requiring the original, unwatermarked image \cite{eggers2001blind}. However, most blind watermarking methods leverage frequency domain features~\cite{cox2007digital}, operating by altering specific Fourier frequencies of the image. Lukas et al.~\cite{lukas2023leveraging} discussed adaptive attacks on image watermarks by exploiting weaknesses in frequency domain watermarking, including detailed attack optimization strategies to bypass watermarking protections. Traditional multimedia watermarks are unchangeable by design. If attackers become aware of this type of singular watermark, forgery and tampering becomes easy. 

\hspace*{20cm}

We propose a unique watermarking mechanism for generative diffusion models (see Fig.~\ref{fig:1}). Our method hides watermark information deep inside the generative content by altering the initial noise signal used for diffusion-based image generation. 
By infusing noisy watermark features across all the denoising steps of the generative process, our watermark becomes an integral, imperceptible part of the content. We propose reverse diffusion process combined with watermark extractor for effective watermark learning/embedding and  extraction/reconstruction. In our method, the watermark image resides in the diffusion model itself so that all images generated by the model contains the watermark.

We incorporate our blind watermarking mechanism into both mainstream diffusion model types, namely; Denoising Diffusion Implicit Models (DDIMs) \cite{song2020denoising} and Denoising Diffusion Probabilistic Models (DDPMs) \cite{ho2020denoising}, generating images with different watermarks for each model type. An image-based neural network classifier is designed to identify the source generative model based on the reconstructed watermarks.  To that end, we also use different watermarks per generative model to construct a Generative Watermarked Image (GWI) dataset. 
Our contributions are summarized below: 

\begin{itemize}
    \item We propose a unique method for embedding imperceptible watermarks into diffusion-based image generation. The watermark is incorporated at every step of the model training, making it an integral, yet imperceptible part of the generated content. 
    \vspace{1mm}
    \item  We introduce an experimental setup to validate the robustness of watermarked diffusion models. We exploit several image-based attack methods to establish our setup, leveraging more than ten types of image statistical measures for comparison and assess the detection and reconstruction performances.
    \item We propose a watermark classification network to verify watermarked image source models. The classification network can distinguish whether the generated images contain watermarks while also identifying the type of watermark, to enable source tracing and copyright establishment.
    \item We presents the GWI dataset, composed of blind watermarked image samples, evidencing four unique watermarks embedded across two diffusion model architectures (DDIM and DDPM). 

\end{itemize}

\section{Related Work}

\subsection{Diffusion Models}

Inspired by non-equilibrium thermodynamics, Ho et al. proposed the popular denoising diffusion probabilistic (DDPM) generative paradigm \cite{ho2020denoising}, which performed competitively compared to PGGAN \cite{guo2021pggan} on the 256x256 LSUN dataset \cite{yu2015lsun}. Since this development, a considerable amount of research has focused on diffusion models to improve architectures, accelerate sampling speeds, and explore various downstream tasks. Nichol et al.~\cite{nichol2021improved} discovered that learning the variances of the reverse process in DDPMs can significantly reduce the number of sampling steps required. Song et al.~\cite{song2020denoising} extend DDPMs through a class of non-Markovian diffusion processes into denoising diffusion implicit models (DDIMs), yielding higher-quality samples with fewer sampling steps. Subsequent work, Adaptive Diffusion Model(ADM) \cite{li2024adm}, identifies a more effective architecture and achieves state-of-the-art performance compared to other generative models with classifier guidance. Viewing DDPMs as solving differential equations on manifolds, Liu et al.~\cite{liu2022pseudo} proposed pseudo-numerical methods for diffusion models (PNDMs), further enhancing sampling efficiency and generation quality. Beyond unconditional image generation, there is a wealth of \textit{conditional} text-to-image generation literature which leverage the diffusion process. Among these, VQDiffusion, based on a VQ-VAE~\cite{van2017neural}, models the latent space with a conditional variant of DDPMs. The latent diffusion model (LDM) \cite{rombach2022high} exploits a cross-attention mechanism and latent spaces to condition diffusion models on textual inputs.

\subsection{Watermarked Diffusion Models}



Recent research in watermarking technology has primarily focused on optimising robustness, imperceptibility, and adaptability. Existing works have provided diverse approaches to address these challenges. Wu et al.~\cite{wu2021watermarking} introduced a method for embedding watermarks into the outputs of neural networks, protecting the intellectual property of deep learning models. Their approach effectively balances image quality and watermark robustness. Liu et al.~\cite{liu2018blind} proposed a dual watermarking mechanism combining robust and fragile watermarks, achieving simultaneous copyright protection and tampering detection. Zong et al.~\cite{zong2015histogram} utilised histogram shape-based features to design a robust watermarking scheme, particularly effective against geometric attacks such as cropping and random bending. Wang et al.~\cite{wang2024template} employed a template-enhanced network to achieve watermark synchronisation under geometric attacks, improving extraction accuracy. Huang et al.~\cite{huang2023texture} leveraged texture-aware adaptive embedding strategies to optimise watermark robustness and imperceptibility, focusing on textured regions to balance these trade-offs.

Compared to these works, our study introduces a novel strategy that embeds watermarks directly into the diffusion process of generative models. Unlike traditional methods that focus solely on protecting output images, our approach ensures that the watermark is deeply integrated with the generation process, enabling traceability back to the model itself. The proposed framework demonstrates strong model-agnostic adaptability, as it can be applied to various diffusion model architectures such as DDIM and DDPM. Furthermore, experimental results reveal that our method maintains high visual quality, with negligible impact on image fidelity, while achieving robustness against multiple attacks, including compression, rotation, and blurring.

A key innovation of our work lies in its efficient encoding and decoding of watermark features using an autoencoder. This design enhances the precision of watermark extraction, even under substantial attack conditions. Additionally, we contribute to the field by introducing the Generative Watermarking Image (GWI) dataset, which provides a standardised benchmark for evaluating watermarking techniques in generative models. Unlike the template-based synchronisation approach proposed by Wang et al.~\cite{wang2024template}, our method eliminates the reliance on complex template embedding, simplifying implementation and reducing computational overhead. Moreover, by integrating watermarking into the diffusion model's training process, we extend the applicability of watermarking technologies beyond the traditional scope, establishing a robust foundation for protecting generative models.

\begin{figure*}
\centering
  \includegraphics[width=\textwidth]{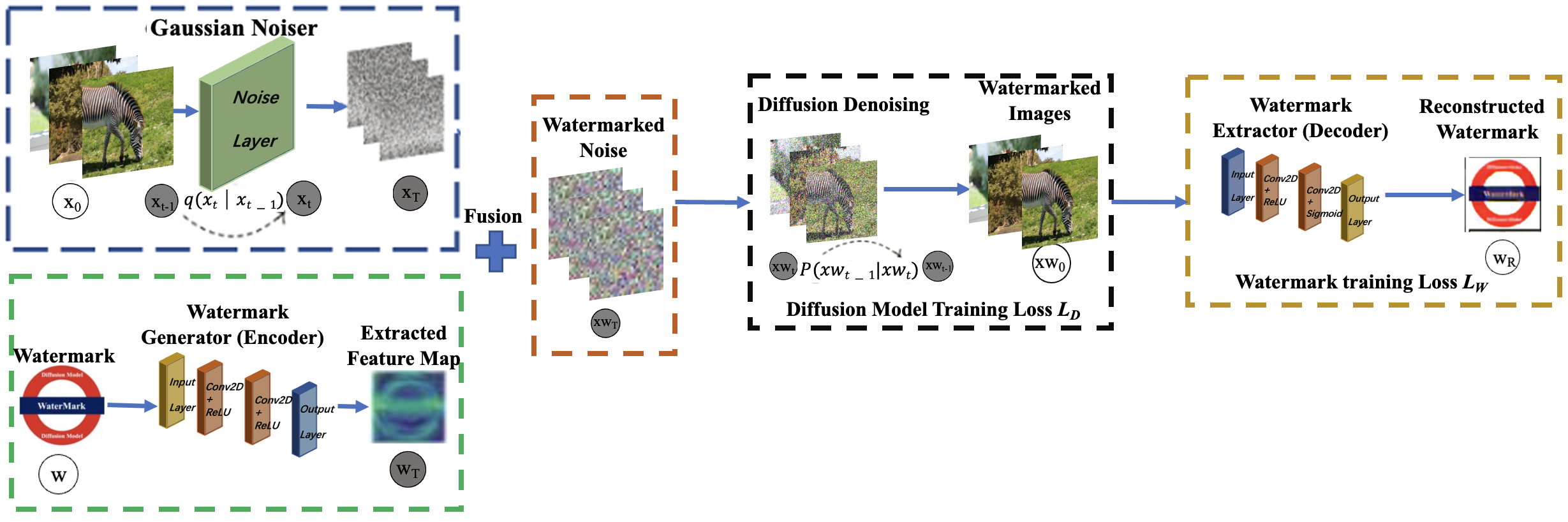}
\caption{\small The Watermark Generator transforms an image-based watermark $w$ into a watermark feature map $w_{T}$ which serves as a branch input to the diffusion model. It is merged with the output $x_{T}$ of the forward diffusion process. The watermarked noise `$xw_T$' serves as a new input for the diffusion denoising phase, which enables embedding watermark features into the diffusion model. The trained diffusion model generates images $xw_{0}$ as usual, but they contain imperceptible watermark information. The Watermark Extractor learns to reconstruct the watermark  $w_R$ from the generated images.}
\label{fig:2}       
\end{figure*}

\section{Method}

We develop an end-to-end method that can embed any image as a blind watermark in diffusion models, see Fig.~\ref{fig:2}. Our method combines the denoising training of diffusion models with autoencoder feature extraction and inverse mapping capabilities. By using image features as branch inputs during the diffusion model training process, the method guides the generative mechanism with blind watermarks. Here, we detail the mathematical principles behind the blind watermarking method and discuss our end-to-end training process. To demonstrate the generality of our technique, we perform end-to-end training on both DDIM and DDPM. 

\subsection{Diffusion Models}
\label{DDIMMethod}

Generative diffusion models `${\mathcal{M}_D}$' \cite{song2020denoising} incrementally add Gaussian noise to input data (blue box in Fig. \ref{fig:2}) and through the reverse denoising process, gain the ability to generate new samples based on learned representations (black box in Fig. \ref{fig:2}). Given a data sample $x_0$, ${\mathcal{M}_D}$ employs a series of Markov transitions \cite{wang2012markov} $q(x_t|x_{t-1})$ to produce a sequence of latent variables $\{x_1, \ldots, x_T\}$, where $T$ denotes the number of diffusion steps. Each step introduces a degree of Gaussian noise into the data, culminating in the final step where the resultant $x_T$ is close to a standard Gaussian distribution. This process can be represented as:

\begin{equation}
\centering
q(x_t | x_{t-1}) = \mathcal{N}(x_t; \sqrt{1 - \beta_t}, \beta_t \mathbf{I}).
\end{equation}

Here, \( \beta_t \) is a hyperparameter that controls the amount of noise added at each step. \( \alpha_t \) is a scaling factor at time step \( t \). According to Markovian properties, we can sample the latent variable \( x_t \) at any time step \( t \) by adding noise in a single step as:

\begin{equation}
q(x_t | x_0) = \mathcal{N}(x_t; \sqrt{\bar{\alpha}_t} x_0, (1 - \bar{\alpha}_t) \mathbf{I}).
\end{equation}

To generate new samples, ${\mathcal{M}_D}$ learns a reverse diffusion process. That is, starting from $x_T$, it produces $x_{t-1},\ldots, x_T$ through a series of inverse Markov transitions $p_0(x_t|x_{t-1})$. Each inverse transition removes a portion of the noise until a clean sample $x_0$ is obtained. This process can be expressed as:
\begin{equation}
p_{\theta}(x_{t-1}|x_t) = \mathcal{N}(x_{t-1}; \mu_{\theta}(x_t, t), \sigma_t^2 \mathbf{I}),
\end{equation}
where, $\mu_{\theta}(x_t, t)$ is a neural network that predicts the denoised sample, and $\sigma_t$ is a hyperparameter controlling the amount of noise removed at each step. During training, ${\mathcal{M}_D}$ minimizes the following loss function:

\begin{equation}
\mathcal{L}_{\text{${\mathcal{M}_D}$}} = \mathbb{E}_{x_0, \epsilon} \left[ \lVert \epsilon - \epsilon_{\theta}(\sqrt{\bar{\alpha}_t} x_0 + \sqrt{1 - \bar{\alpha}_t} \epsilon, t) \rVert^2 \right],
\label{l1} 
\end{equation}
where $\epsilon$ is standard Gaussian noise, and {$\epsilon_{\theta}$ is a neural network} 
for predicting the noise added.

\subsection{Watermark Autoencoder}

Our watermark encoder is designed to embed watermark information into noisy representations of image data. The encoder `$E(\cdot)$' maps the watermark image $w$ to a low-dimensional latent representation $w_T$, 

\begin{equation}
w_T = E(w).
\end{equation}
We combine the latent representation from the watermark autoencoder \( w_T \) with the Gaussian noise representation \( x_T \) generated by ${\mathcal{M}_D}$ to obtain the watermarked noise as visualized in Fig.~\ref{fig:2}: 

\begin{equation}
xw_T = w_T + x_T.
\end{equation}

We then input the watermarked noise $xw_T$ into the reverse ${\mathcal{M}_D}$ diffusion process to generate the watermarked image $xw$:

\begin{equation}
\ xw = {\mathcal{M}_D}_r (xw_T).
\end{equation}

\begin{figure}
  \centering
  \includegraphics[width=\columnwidth]{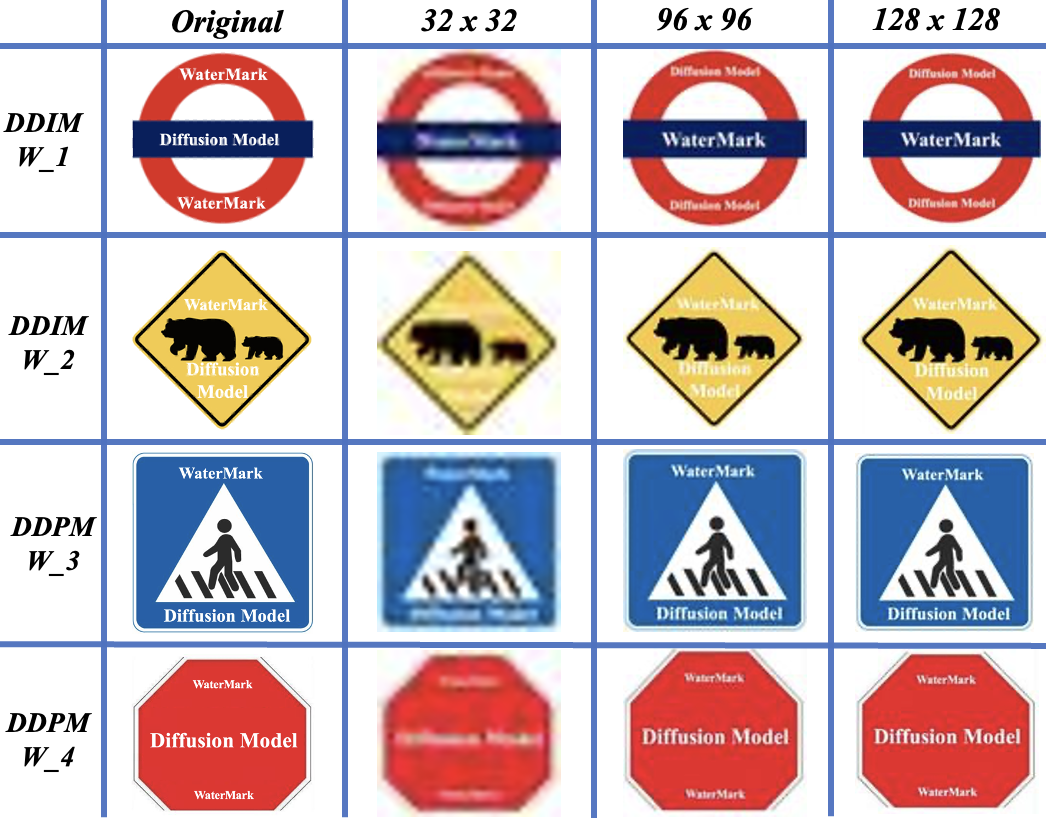}
\caption{Our blind watermarking mechanism is incorporated into DDIM and DDPM, generating images with two unique watermarks per model. Each model was trained and analyzed at three different resolutions, with the watermarks also adjusted to the corresponding resolutions for training. }
\label{resolution}       
\end{figure}

We incorporate our blind watermarking mechanism into DDIM and DDPM, generating images with two unique watermarks per model as shown in Fig. \ref{resolution}. Each model is trained and analyzed at three different resolutions (32$\times$32, 96$\times$96, 128$\times$128), with the watermarks also adjusted to the corresponding resolutions for training. This ensures that the watermarking process is not affected by image resolution.

\subsection{Watermarked Model Architecture}

In this section, we delve into the architecture of our model's watermark encoder and decoder. The watermark encoder is designed to embed watermark information into the latent representation of the image data. Specifically, the encoder, denoted as \( E \), consists of a series of convolutional layers followed by activation functions and pooling layers to progressively reduce the spatial dimensions and generate a compact watermark representation, \( w_T = E(w) \), where \( w \) is the input watermark image. This compact representation is then combined with the Gaussian noise representation generated by the diffusion model to obtain the watermarked noise, as visualized in Fig. 2. The reverse diffusion process is then used to generate the watermarked image \( x_w \), ensuring that the watermark is embedded at multiple levels of the diffusion process.

The decoder `\( D(\cdot) \)', is responsible for extracting and reconstructing the watermark from the generated image. 
The decoder uses upsampling and deconvolution layers to transform the watermarked noise back into a perceptible watermark image. This process is optimized to retain watermark information while ensuring minimal impact on the visual quality of the generated image. Our model's end-to-end training framework integrates the loss functions of the diffusion model and the autoencoder. The diffusion model’s loss ensures high-quality image generation, while the autoencoder’s loss focuses on watermark embedding and extraction accuracy. 

The diffusion models employed in this work are predicated on the U-Net architecture \cite{rombach2022high}, encompassing an encoder, a decoder, and an attention mechanism. The encoder comprises numerous downsampling blocks, residual blocks, and attention blocks. Correspondingly, the decoder is structured with multiple upsampling, residual, and attention blocks. Furthermore, an intermediate `mid-' block is integrated between the encoder and decoder to account for global information. Traditional autoencoders typically consist of reduction layers to produce compressed representations, with this process being reversed in the decoder. As visualized by the green and yellow boxes in Fig.~\ref{fig:2}, we maintain a consistent size across all convolutional layers of our autoencoder (including the encoder and decoder). This approach aims to retain more details and nuances of the input data, potentially enhancing the model's performance in tasks where such finer details are important for generation. 

\subsection{End-to-End Training}

Our model employs an end-to-end training approach that intuitively combines diffusion model features with auto-encoder features. During training, the model learns to generate images with high visual quality and with embedded imperceptible watermarks. Our final loss function $L_{\text{Total}}$ is derived from both the diffusion model and autoencoder loss functions. Given losses derived in Eq.~(\ref{l1}) and later in Eq.~(\ref{l2}), $L_{\text{Total}}$ is  defined as:

\begin{equation}
L_{\text{Total}} = L_{\text{${\mathcal{M}_D}$}} + L_{\text{w}}.
\end{equation}

Specifically, the diffusion model component of the loss $L_{\text{${\mathcal{M}_D}$}}$ ensures the quality of image generation and the denoising accuracy. In contrast, the autoencoder's loss $L_{\text{w}}$  optimizes the watermark embedding and extraction process. Accounting for both loss functions enables the model to focus on generating and protecting image content during training, achieving the dual objectives of high-quality image generation and watermark protection. Our model retains a high generative quality while ensuring the establishment of model copyright.

\begin{figure*}
    \centering
    \includegraphics[width=\textwidth]{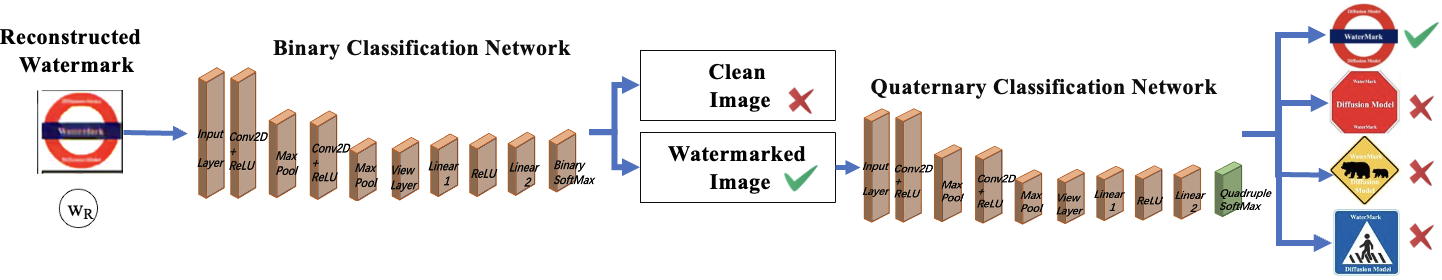}
\caption{ The Classification Process for Watermark Detection and Identification. The input first passes through a Binary classification network to determine whether it contains a blind watermark. It then goes through a Quaternary classification network to identify the type of watermark. The two classification networks differ only in the final softmax layer.}
\label{two}       
\end{figure*}

\subsection{Watermark reconstruction and classification}

\label{classifcationNet}

For watermark reconstruction, we first use the forward diffusion process  to transform the watermarked image $xw$ into the watermarked noise $\tilde{xw}_T$:

\begin{equation}
\tilde{xw}_T = {\mathcal{M}_D}_f (xw).
\end{equation}

Then, we use the decoder $D$ of the watermark autoencoder to reconstruct the watermarked image $w_R$ from the watermarked noise as:
\begin{equation}
\ w_R = D(\tilde{xw_T}).
\end{equation}

During the training process, the watermark autoencoder minimizes the following reconstruction loss:

\begin{equation}
L_{\text{w}} = \| w - w_R\|^2.
\label{l2}
\end{equation}

As shown in Fig. \ref{two}, we construct two classification networks to identify the presence and type of human-imperceptible watermarks embedded in generated images. The first network is designed to detect whether an image contains a watermark. The second network identifies the specific type of watermark embedded.

\subsubsection{Watermark Presence Detection Network}

The Watermark Presence Detection Network determines if a watermark is embedded within a given image. This network processes the input image through a series of convolutional layers, each layer employing small kernels to capture detailed, local image features. The convolutional layers are followed by ReLU activation functions such  to introduce non-linearity and pooling layers that reduce the spatial dimensions of the feature maps, thus condensing the essential information while making the network more efficient.

After sufficient feature extraction through these layers, the network flattens the resulting feature maps and passes them through fully-connected layers. These dense layers integrate the learned features into a compact representation, forming a final, binary classification layer. We utilize the softmax function for binary watermark presence detection.


\begin{table*}[h]
     \scriptsize
    \centering
    \renewcommand\arraystretch{1.3}
    \setlength{\tabcolsep}{1mm}{
    \caption{We select 10 image statistics (see Section~\ref{sta}) based on 3 aspects (texture, edge, and frequency) measured on generated images at 3 resolutions ((32$\times$32, 96$\times$96 and 128$\times$128). The results indicate a less than 2\% overall change rate between watermarked and unwatermarked images, across the 10 statistical measures. ``Difference'' refers to the percentage difference in data. The data here is derived from the average of 300 random images from the GWI dataset.}
    \label{Statistics}
    \resizebox{\linewidth}{!}{%
    \begin{tabular}{l|c|c|c|c|c|c|c|c|c|c|c}
    \cline{2-12}
     & GLCM  & GLCM  & Canny  & Variance Blur & Mean  & Edge  & Entropy & Sharpness & Saturation & Texture   &  Image \\ 
     & Contrast & Energy & Edge & Measure & spectrum & Histogram & \cite{tsai2008information} & \cite{de2013image}  & \cite{de1996naturalness} & Strength  &  Resolution \\
     & \cite{sebastian2012gray} & \cite{sebastian2012gray} & \cite{tahmid2017density}& \cite{he2005laplacian}&\cite{frank2020leveraging}&\cite{chan2018convex}& & & & \cite{freitas2018application}  &  \\
     
    \cline{1-12}
    Watermarked & 467.90 & 0.027 & 61.91 & 2323.27 & 115.27 & 4.00 & 7.22 & 10611.32  & 81.04 & 3.65  & \\
    Clean       & 470.65 & 0.027 & 62.57 & 2345.83 & 115.38 & 4.00 & 7.23 & 10541.17  & 83.47 & 3.63  & 32 $\times$ 32 \\
    Difference  & 0.58\% & 0.000\% & 1.06\% & 0.97\% & 0.1\% & 0.00\% & 0.07\% & 0.66\%  & 3.01\% & 0.58\%   & \\
    \hline
    \cline{1-12}
    Watermarked & 213.18 & 0.025 & 32.97 & 1878.25 & 131.18 & 36.00 & 7.24 & 12474.23 & 51.06 & 3.71   & \\
    Clean       & 220.67 & 0.022 & 34.03 & 1888.01 & 131.36 & 36.00 & 7.27 & 12346.23 & 53.15 & 3.67  & 96 $\times$ 96\\
    Difference  & 3.39\% & 13.63\% & 3.12\% & 0.51\% & 0.14\% & 0.00\% & 0.31\% & 1.03\%  & 5.10\% & 0.95\%   & \\
    \hline
    \cline{1-12}
    Watermarked & 188.32 & 0.019 & 128.32 & 1723.83 & 120.38 & 64.00 & 7.30 & 13927.23 & 43.23 & 3.81  & \\
    Clean       & 192.32 & 0.020 & 132.35 & 1792.32 & 122.38 & 64.00 & 7.31 & 13899.38 & 45.23 & 3.82  & 128 $\times$ 128\\
    Difference  & 2.12\% & 5.26\% & 3.13\% & 3.97\% & 1.66\% & 0.00\% & 0.11\% & 0.20\%  & 4.63\% & 0.23\%  & \\
    \hline
    \end{tabular}}}
\end{table*}

\subsubsection{Watermark Type Identification Network}

The Watermark Type Identification Network builds upon the foundational architecture of the first network but is designed for a multi-class classification task. This network begins similarly, with convolutional layers that progressively extract more abstract features from the input image. However, to handle the complexity of identifying multiple watermark types, this network is deeper, with more convolutional layers that allow for finer feature extraction.
The feature maps generated by these layers are then passed through fully connected layers that are more complex than those in the presence detection network. These layers are responsible for distilling the information into a form that can be used to distinguish between different watermark types. The final classification layer uses a softmax function to output a probability distribution across multiple classes, each corresponding to a specific watermark type.


\begin{table*}[h]
    \centering
    \scriptsize
    \renewcommand\arraystretch{1.5}
    \setlength{\tabcolsep}{1mm}{
    \caption{Inception Score (IS)~\cite{salimans2016improved} and Frechet Inception Distance (FID)~\cite{heusel2017gans} values calculated at 3 resolutions (32$\times$32, 96$\times$96, and 128$\times$128). Each resolution dataset includes 200 images and 2 categories: (i) watermarked and (ii) unwatermarked images. Small percentage ``Difference'' between watermarked and unwatermarked images shows the robustness of our method.}
    \label{tab:rotation}
    \resizebox{0.9\linewidth}{!}{%
    \begin{tabular}{l|ccc|ccc|ccc}
    \cline{2-10}
    & \multicolumn{3}{c|}{32 $\times$ 32} & \multicolumn{3}{c|}{96 $\times$ 96} & \multicolumn{3}{c}{128 $\times$ 128} \\  
    \cline{2-10}
    & Clean  & Watermarked& Difference &Clean &Watermarked& Difference& Clean &Watermarked& Difference\\
    \cline{1-10}
    IS & 1.98 & 2.02 & 1.64\% & 2.61&2.44&6.84\%&2.24&2.01&9.98\% \\
    \cline{1-10}
    FID & 1046.52 &967.63 & 7.54\% &875.39 & 856.38 & 2.17\%&811.35&782.43& 3.56\%\\
    \cline{1-10}
    \end{tabular}}}
    \vspace{-0mm}
\end{table*}

\section{GWI Dataset}
\label{dataset}

We have constructed the Generative Watermarked Image (GWI) dataset, containing four types of watermarks. The GWI  dataset is specifically constructed to evaluate the effectiveness and robustness of our watermarking technique on diffusion models. This dataset is designed to provide a diverse set of watermarked images across different resolutions and watermark types, ensuring comprehensive testing of watermark embedding, detection, and extraction processes. The GWI dataset contains images generated using DDIM and DDPM generative models, embedded with distinct watermarks to facilitate source tracing and copyright verification.

\subsection{Dataset Composition}

The GWI dataset comprises three main resolution categories: 32$\times$32, 96$\times$96, and 128$\times$128. Four unique watermarks are embedded for each resolution, resulting in four distinct watermark types. Each resolution category contains a total of 400 images, with 100 images per watermark type. In addition to the watermarked images, an equivalent set of non-watermarked (clean) images is included for comparison purposes, bringing the total number of images in the GWI dataset to 2400.

\subsection{Purpose and Utility}

The GWI dataset is a valuable resource for testing and validating the robustness of watermarking techniques in generative models. By providing a controlled set of watermarked and clean images across multiple resolutions, the dataset allows for detailed analysis of watermark imperceptibility, robustness against image transformations, and the effectiveness of watermark extraction methods. This dataset serves as a benchmark for future research in the field of generative model watermarking.

\section{Experiment results and discussion}

To conduct our experiments and validate the generalizability of our method, we considered three different datasets with varied image resolutions (32$\times$32, 96$\times$96 and, 128$\times$128) for training.
For the 32$\times$32 resolution, we utilized the CIFAR-10 dataset~\cite{cifar10dataset}, which contains 60,000 images, divided into 10 labelled classes, i.e., 6,000 images per class. 48,000 images were used for training, and 12,000 images were used for testing. Due to its relatively small data size and diverse class labels, CIFAR-10 continues to be a widely popular dataset in the computer vision domain. For the 96$\times$96 resolution, we selected the STL-10 dataset~\cite{stl10dataset}. This  dataset includes 10 classes, the same as those in CIFAR-10. The dataset contains 13,000 color images, with 8,000 used for training and 5,000 for testing. For the 128$\times$128 resolution, we chose the Oxford-IIIT Pet Dataset~\cite{oxfordpetsdataset}, which contains around 7,000 images covering 37 different breeds of dogs and cats, with about 200 images per breed. 4,000 images were used for training, and 3,000 images were used for testing.
We also design a novel experimental procedure to test watermark robustness, using various watermark attack methods for our evaluations. To aid in our discussion and provide added transparency, we also present some worst-case examples of our watermark embedding and image generation method.

\subsection{Generative Quality}
\label{sta}

Several works highlight the importance of texture, edges, and frequency for image detection~\cite {do2005contourlet,rossler2019faceforensics++,cozzolino2017recasting,afchar2018mesonet}. We designed an experimental process, outlining ten image statistical measures for watermarked image analysis. These measures were chosen based on their ability to evaluate texture, edges, and frequency information. Table~\ref{Statistics} reports the results for ten statistical measures, varying the image resolution. These measures are extracted for the generated images during our experiments. The experimental results presented in Table~\ref{Statistics} report the mean values based on the resolution-depend- ent image set (as first described in section \ref{dataset}).

We present the experimental results for the watermarked and clean images and compare their differences in Table~\ref{Statistics}. Almost all differences are less than 2$\%$, except for GLCM Energy \cite{tsai2008information} at 96$\times$96 and 128$\times$128 image resolutions, where the difference is $>5\%$. GLCM Energy is a texture measure, a statistical method of examining texture that considers the spatial relationship of pixels. It  provides a measure of textural uniformity or smoothness in an image. Higher energy values indicate a lower texture or more homogeneity, meaning the pixel pairs have less variation and are more alike. 

The higher difference in GLCM Energy indicates that the watermark embedding process slightly affects the textural uniformity of the images at higher resolutions. This suggests that while the watermark remains visually imperceptible, it introduces minor variations in pixel relationships that are more detectable at higher resolutions. This can be attributed to the increased detail and pixel density available at higher resolutions, making even small texture changes more noticeable. Despite these differences, the overall impact on image quality remains minimal, demonstrating the robustness of our watermarking method. In addition, GLCM Energy values are typically very small. However,
these minor changes can result in large percentage differences.


We also used Inception Score (IS)~\cite{salimans2016improved} and Frechet Inception Distance (FID)~\cite{heusel2017gans} to compare the quality of watermarked vs. clean image sets. IS evaluates the clarity and diversity of generated samples using a pre-trained neural network, while FID measures image quality by comparing the distribution parameters (mean and covariance) of generated and real data embeddings. As part of our analysis, we compare the differences of IS and FID scores on watermarked and clean images. Table \ref{tab:rotation} shows that the differences at three resolutions were below 5$\%$, indicating that our model produces blind watermarked images with very low deviation from the clean images. We observe that the deviation increases w.r.t the image resolution. We believe this is because higher-resolution images contain more detailed information, which might make the watermark's impact on perceived image quality more noticeable. This finding underscores the effectiveness of our model in maintaining image quality, even at higher resolutions, thus confirming its potential and applicability in real-world scenarios.


\begin{figure*}
    \centering
    \includegraphics[width=0.95\textwidth]{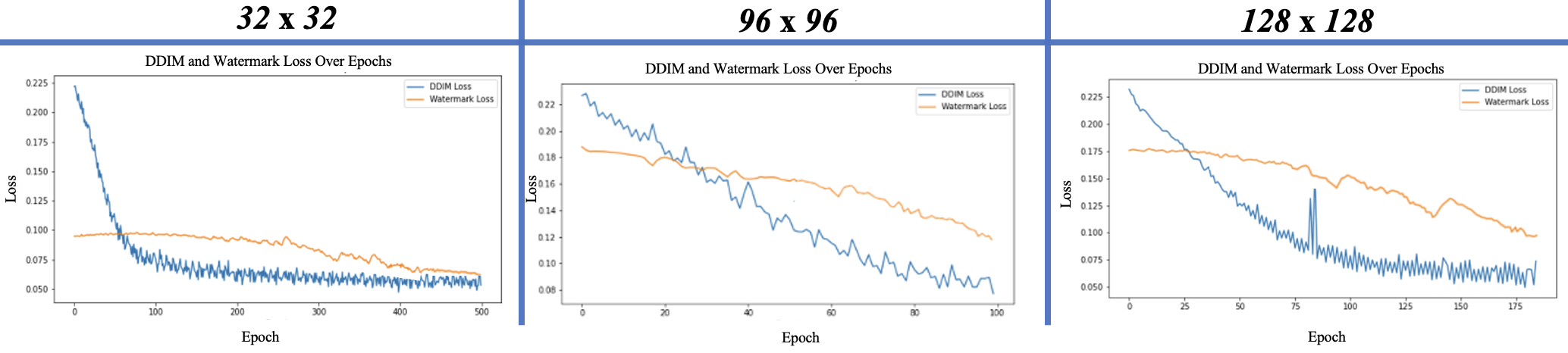}
\caption{Loss analysis of the end-to-end training process. The Y-axis represents the change in loss, while the X-axis represents the increase in epochs. The figure illustrates the loss curves for the diffusion model $L_{\text{D}}$ and autoencoder $L_{\text{W}}$.
We asses our performances across three image resolutions. We can observe that $L_{\text{D}}$ (blue curve) converges rapidly and soon levels with $L_{\text{W}}$ (yellow curve). }
\label{Loss}       
\end{figure*}

\begin{figure*}
\centering
  \includegraphics[width=0.95\textwidth]{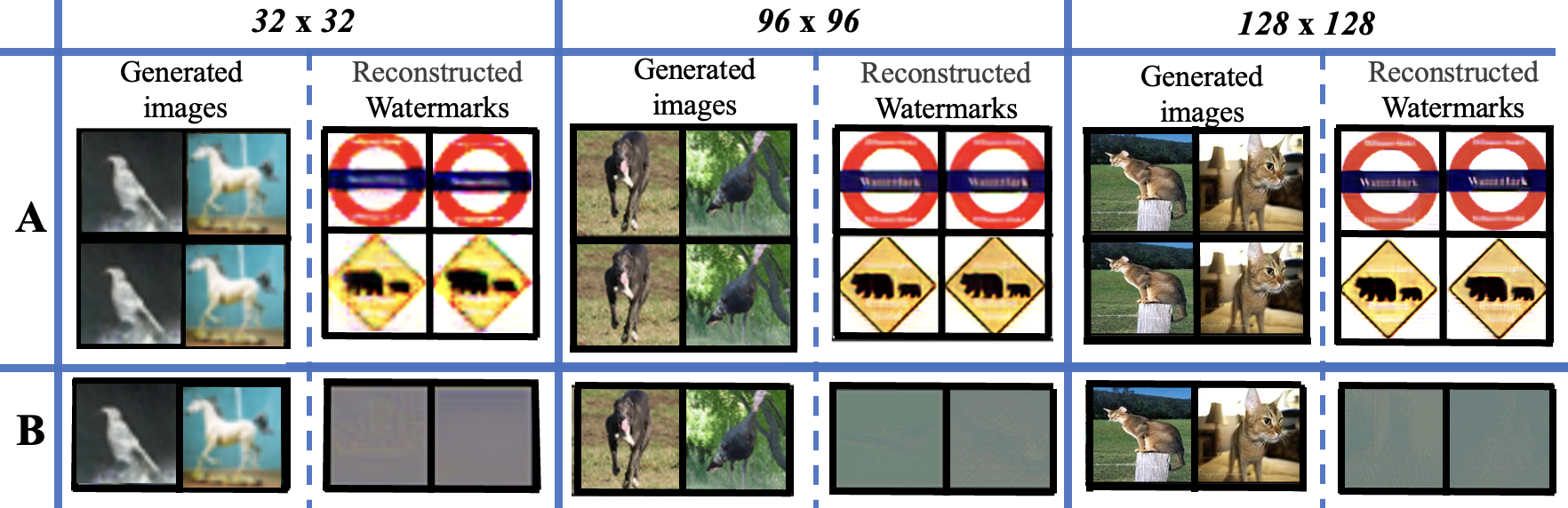}
\caption{Visual representation of blind Watermarked-DDIM-Generated images at three resolutions. Row A displays the generated images with embedded watermarks and the reconstructed watermarks, highlighting the two unique DDIM watermark types. Row B presents the attempted watermark reconstruction results from clean images, demonstrating only noise-like representations are output. }
\label{DDIMR}       
\end{figure*}


\subsection{Loss Function Analysis and Weight Adjustment}

Figure~\ref{Loss} shows the loss curves at three different image resolutions. We observe that the convergence rate of the  DDIM loss $L_{\text{DDIM}}$ is significantly faster than that of the watermark loss $L_{\text{w}}$. A rapidly converging loss function may reach a lower error level early in training but subsequently contributes less to improving model performance. To balance the impact of both loss terms on the training dynamics of the model and promote a more balanced optimization of performance throughout the training process, we assigned a higher weight to $L_{\text{DDIM}}$. This weight adjustment is intended to ensure that even after $L_{\text{DDIM}}$ converges quickly, the model can improve by optimizing $L_{\text{w}}$. 
This weight adjustment was carefully determined through a series of experiments designed to balance the dual objectives of our model: maintaining high image quality and ensuring robust watermark embedding. 
Through empirical analyses, we identified specific weight combinations for training optimization, allowing the model to generalize well across different datasets and image resolutions. This dynamic weighting approach promotes a more balanced performance optimization throughout the training process, ensuring that the model does not prematurely converge on image quality at the expense of watermark robustness.

\subsection{Watermark Reconstruction and Classification } 
\label{reconstraction}

\begin{figure*}
\centering
  \includegraphics[width=0.95\textwidth]{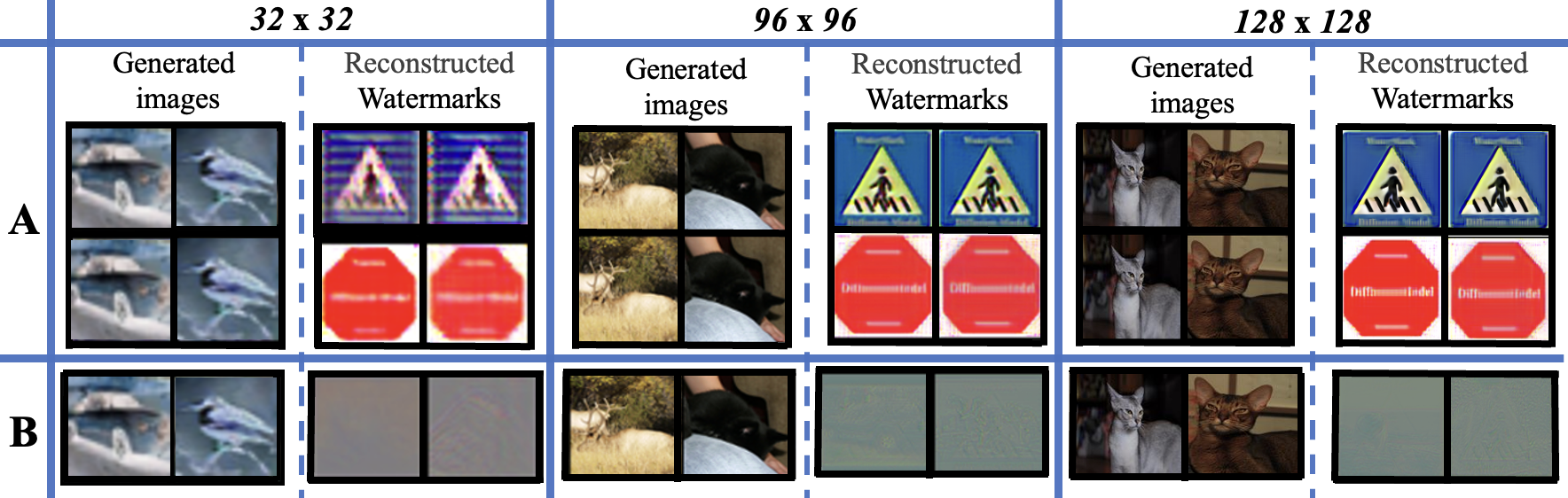}
\caption{Visual representation of blind Watermarked-DDPM-Generated images at three different resolutions. The top row displays the DDPM-generated images with embedded watermarks and the reconstructed watermarks. The bottom row presents the attempted watermark reconstruction results from clean images, featuring only noise-like representations are output.  }
\label{DDPMR}       
\end{figure*}


\begin{figure}
\centering
  \includegraphics[width=0.45\textwidth]{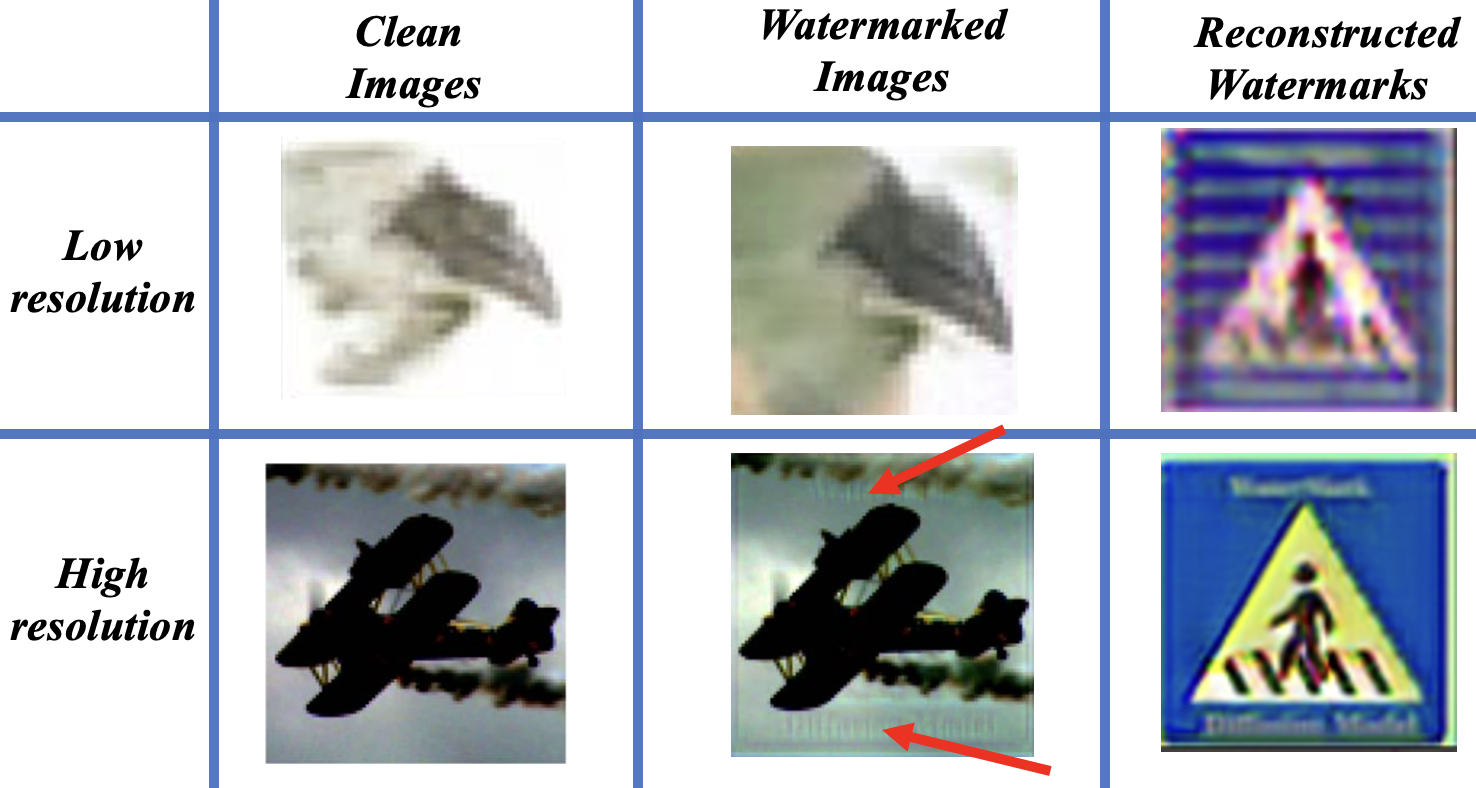}
\caption{Comparison of watermark embedding and reconstruction across different resolutions. The left column shows the original clean images, the middle column displays the images with embedded watermarks, and the right column presents the reconstructed watermarks. Both low and high-resolution images are compared, demonstrating the impact of resolution on watermark visibility and reconstruction accuracy. }
\label{Bad}       
\end{figure}

\begin{table*}[h]
    \centering
    \scriptsize
    \renewcommand\arraystretch{1.1}
    \setlength{\tabcolsep}{1mm}{
    \caption{Classification results of Watermarked Images under various attacks. The classification network distinguishes whether an image contains a watermark and then identifies the type of watermark. The dataset uses 96$\times$96 resolution images, which derived from the Generative Watermarked Image (GWI) dataset .}
    \label{tab:attacks}
    \resizebox{\linewidth}{!}{%
    \begin{tabular}{l|c|c|c|c|c}
    \cline{3-6}
     \multicolumn{2}{c|}{} & \multicolumn{4}{c}{Attack Type} \\
    \cline{2-6}
    & No Attack & Rotation & Low-level Blurring  & Texture Reduction & Image Compression \\
    \cline{1-6}
     Watermark Presence & 100.00\% &88.46$\%$  & 96.15$\%$ & 93.59$\%$ &93.85$\%$ \\
     \cline{1-6}
     Watermark Type     & 82.51\% &75.96$\%$  & 81.97$\%$ & 75.79$\%$ &79.02$\%$ \\
    \cline{1-6}
    \end{tabular}}}
    \vspace{-0mm}
\end{table*}

\subsubsection{Analysis of Watermark Reconstruction}

Results discussed in Sec.~\ref{sta} show that it is hard to distinguish between the watermarked and clean images. The ability to reconstruct and identify watermarks is a key task to ensure copyright protection. Through Fig.~\ref{resolution}, we observe that the quality of the reconstructed watermark improves as the image resolution increases - becoming more closely aligned with the original image for higher-resolution cases. Nevertheless, the watermark features are also retained well at lower resolutions. The adaptive and robust nature of our watermarking technique is demonstrated through the improvement in watermark reconstruction quality w.r.t. image resolution.


As illustrated in Fig. \ref{DDIMR} and Fig. \ref{DDPMR}, we present visual examples of watermark reconstruction across different resolutions. The top row in each section displays the watermarked images, while the corresponding reconstructed watermarks are shown in the adjacent columns. The bottom row in each section presents the clean images without watermarks for comparison. 
For each resolution, the first column shows generated images with embedded watermarks and the second column shows the \textit{reconstructed} watermarks.
The quality of the reconstructed watermarks improves w.r.t. image resolution, demonstrating the scalability of our approach. Higher resolutions retain more detailed information, contributing to more accurate reconstruction.

Watermark capacity is considered an important property. Using an entire image as the watermark, we can easily transmit information such as the copyright holder organization/name. This allows for easy contact with the rightful owner for rights protection.  Our method of embedding watermarks is more practical and meaningful than traditional single-frequency domain watermarks. Traditional methods that operate in the frequency domain are easily targeted by attackers who understand the specific frequencies that have been altered. In contrast, our spatial domain approach disperses the watermark information throughout the image, enhancing its resilience against such attacks.

The image statistical measures and IS/FID results reported above indicate a minimal difference between watermarked and clean images as intended. Most of the generated results are indistinguishable from the human vision system.  We present a representative ``poorly generated result" in Fig.~\ref{Bad}, comparing low and high-resolution images. In low-resolution images, watermarked images can show slight color variations when compared to the clean image. These variations are primarily due to minor changes in pixel values introduced during the watermark embedding process. The overall image structure remains largely intact and the perceptual quality is maintained, despite the color differences. In contrast, high-resolution images reveal additional artifacts caused by the watermark embedding. Beyond the color differences, the high-resolution watermarked image exhibits slight ``ghosting'' artifacts, as indicated by the red arrows in Fig. \ref{Bad}. These artifacts are more pronounced due to the increased level of detail in high-resolution images, making any distortions or imperfections more noticeable. Nevertheless, the image perceptual quality is generally not significantly degraded, and the watermark reconstruction is effective.

\subsubsection{Watermark Reconstruction and Classification}

After successfully reconstructing watermarks, classification accuracy is another crucial aspect to consider. High classification accuracy is essential to ensure that watermarks serve their purpose as model signatures, enabling rights holders to track and manage their content effectively. As per our `No-Attack' results reported in Table \ref{tab:attacks}, our model achieves a 100$\%$ accuracy rate in detecting the presence of blind watermarks in generated images. Furthermore, it correctly classifies the type of watermark with an accuracy of 82.51$\%$.

The No-Attack results align with the design philosophy of our model: the reconstruction and traceability of watermarks, which is crucial for copyright protection. By employing an encoder/decoder architecture that does not alter the size of the watermark, our method ensures that an almost lossless watermark is integrated into generated images and allows for the subsequent accurate reconstruction of the watermark during the inverse mapping process. This high level of performance provides a robust guarantee for copyright protection, enabling tracing of the model's origin. Such capabilities are crucial for ensuring the integrity and authenticity of digital media.

\subsection{ Watermark Attack Performance}

\subsubsection{Overview of Attack Scenarios}

To ensure that our watermarking method is sufficiently robust, we evaluate our method's resilience to various common watermark attacks. We deploy these attacks to obfuscate or remove the watermark. We posit that only attacks which do not affect the visual quality of the generated images are considered effective. In other words, the attacked images should maintain the same visual quality as those that are not attacked, with the attack targeting only the watermark. For example, although adding noise to an image can significantly decrease watermark classification accuracy, the attacked image exhibits noticeable tampering and thus, makes the attack human-perceptible.  Therefore, we consider attacks such as rotation, low-level blurring, texture reduction, and low-level image compression the most widespread. 

\subsubsection{Impact of Attacks on Watermark Classification}

We applied four different types of attacks to our watermarked images to evaluate the robustness of the watermarks. Following these attacks, we analyzed the attacked images using our watermark classification networks. The performance of each classification network under these different attack scenarios is summarized in Table \ref{tab:attacks}. The average accuracy of “watermark presence”  after the attacks decreased by $\approx7\%$. In addition, the average accuracy of “Watermark Type”  after the attacks decreased by about 4.3$\%$. 

This decrease in accuracy can be attributed to the fact that these attacks introduce distortions that affect the visibility and distinctiveness of the embedded watermarks. For instance, rotation can alter the geometric alignment of the watermark within the image, making it harder for detection algorithms to recognize the pattern \cite{wang2004image}. Low-level blurring reduces the clarity of the image, thereby diminishing the fine details crucial for accurately identifying watermarks \cite{lee2006robust}. Texture reduction attacks remove essential textural information from the image, which can obscure the embedded watermark features \cite{zhang2017study}. Image compression, particularly lossy compression methods like JPEG, introduces compression artifacts that can interfere with the embedded watermark, affecting its detection and classification \cite{cox2007digital}. Despite introducing these distortions, our method maintained a relatively high classification accuracy, demonstrating its robustness and practical utility.

\section{Conclusion}

Given the rising concerns of copyright protection in generated visual content, we developed a novel method for embedding hidden watermarks in diffusion models. This method combines autoencoders with diffusion models to hide watermark information while preserving image quality. Integrating watermark features with noise at every denoising step during training improves watermark imperceptibility. Our method's robustness was validated through extensive tests, incorporating a series of watermark attacks and evaluations of various image statistical measurements. Additionally, we present a unique, Generative Watermarked Image (GWI) dataset comprising 1,200 watermarked images from DDIM and DDPM models, with four unique watermarks and three image resolution subsets. Our method offers a scalable and practical solution for maintaining the authenticity and ownership of generation content across various platforms and devices, providing a significant step forward in blind watermarking and generative models.



\section{Acknowledgements}\label{S7}

This research was supported by National Intelligence and Security Discovery Research Grants (project\# NS2 20100007), funded by the Department of Defence Australia. Dr. Naveed Akhtar is a recipient of the Australian Research Council Discovery Early Career Researcher Award (project number DE230101058) funded by the Australian Government. Professor Ajmal Mian is the recipient of an Australian Research Council Future Fellowship Award (project number FT210100268) funded by the Australian Government.

\section{Competing Interest and Conflict }
All authors certify that they have no affiliations with or involvement in any organization or entity with any financial or non-financial interest in the subject matter discussed in this manuscript

{\small
\bibliographystyle{ieee_fullname}
\bibliography{egbib}

}

\section{Biography Section}

\vspace{-10mm}
\begin{IEEEbiography}[{\includegraphics[width=1in,height=1.25in,clip,keepaspectratio]{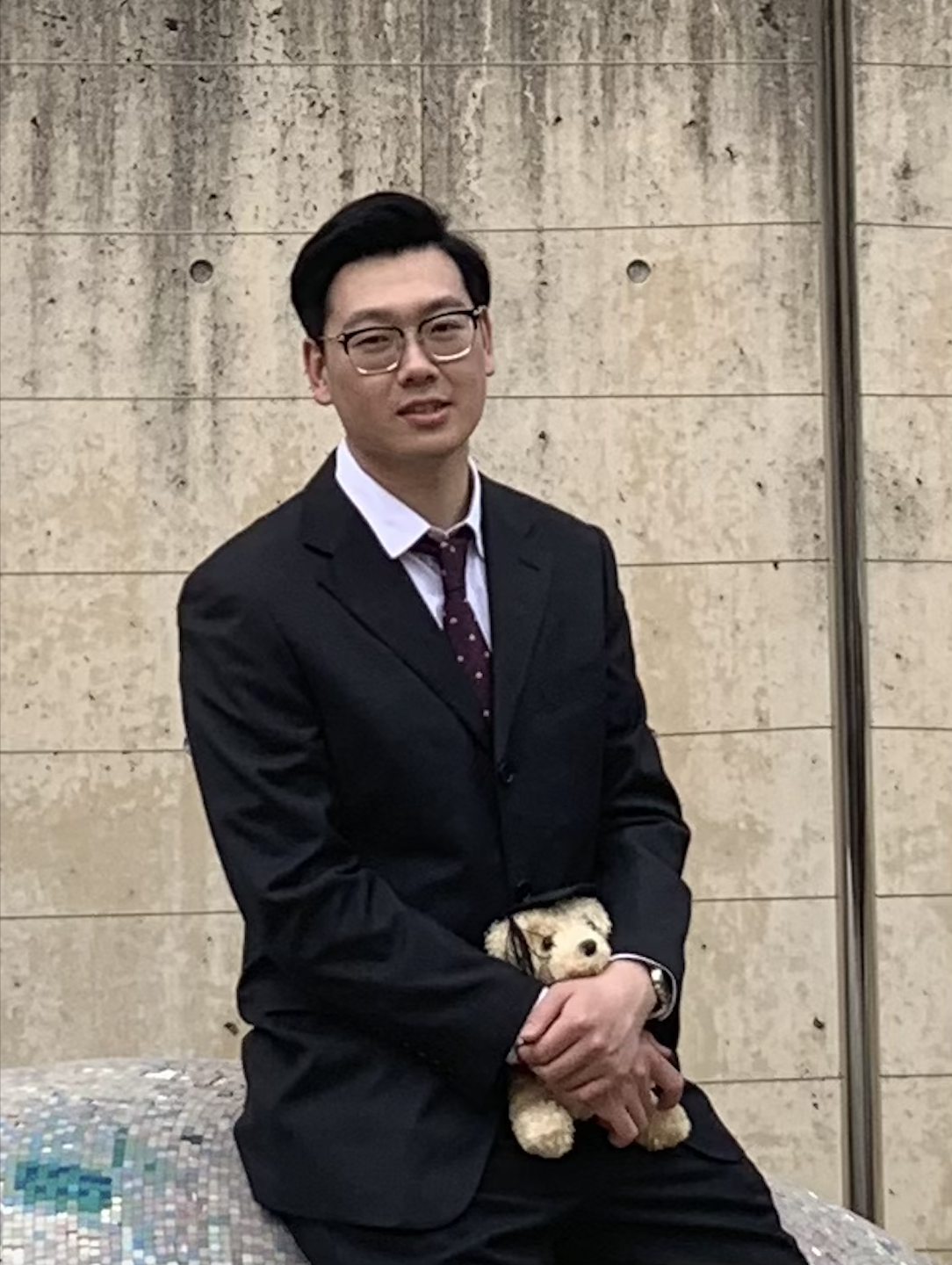}}]{Yunzhuo Chen}
Yunzhuo Chen received the B.Eng. degree in software engineering from the University of Western Australia, Perth, Australia, in 2019, and the M.Eng. degree in software engineering with high distinction from the same university in 2021. He was awarded the Ph.D. scholarship by the University of Western Australia in 2021, where he focuses on research in artificial intelligence, machine learning, diffusion models, network information dissemination, security, and copyright protection. Dr. Chen has presented his research at international conferences in 2022, 2023, and 2024, garnering significant interest from fellow researchers.
\end{IEEEbiography}

\vspace{-10mm}

\begin{IEEEbiography}[{\includegraphics[width=1in,height=1.25in,clip,keepaspectratio]{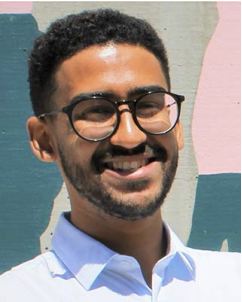}}]{Jordan Vice}
Jordan Vice received the B.Eng. (first class Hons) degree in mechatronic engineering from Curtin University, Perth, Australia, in 2019. He received the Ph.D. degree in mechatronic engineering from the same university in 2022. His research interests include applied artificial intelligence, explainable AI (XAI), machine learning, real-time, multimodal assessment of affective states, and affective computing. Dr. Vice received the 2019 Proxima Consulting Prize for Most Outstanding Final Year Project in mechatronic engineering. His works attracted media attention in 2019, 2020, and 2022.
\end{IEEEbiography}

\vspace{-10mm}
\begin{IEEEbiography}[{\includegraphics[width=1in,height=1.25in,clip,keepaspectratio]{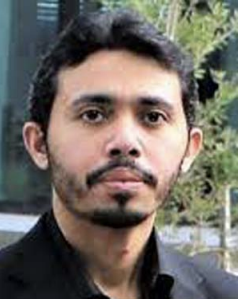}}]{Naveed Akhtar}
(Member, IEEE) received the Ph.D. degree in computer science from the University of Western Australia (UWA). He is currently a Senior Lecturer with the University of Melbourne. His research interests include adversarial machine learning, explainable artificial intelligence, and computer vision applications. He is a recipient of two prestigious Australian Research Fellowships, including the Australian Research Council’s Early Career Researcher Award (ARC DECRA). He is among the top 1 $\%$ scientists in his field for the year 2022. He is a Universal Scientific Education and Research Network 2023 Laureate in Formal Sciences and a recipient of the Google Research Scholar Program Award, in 2023. He was a finalist for Western Australia’s Early Career Scientist of the Year in 2021. He is also an ACM Distinguished Speaker and serves as an associate editor for IEEE Transactions on Neural Network and Learning Systems. He has also served as an area chair for the IEEE Conf. on Computer Vision and Pattern Recognition, European Conference on Computer Vision, and IEEE Winter Conf. on Applications of Computer Vision.
\end{IEEEbiography}
\vspace{-10mm}
\begin{IEEEbiography}[{\includegraphics[width=1in,height=1.25in,clip,keepaspectratio]{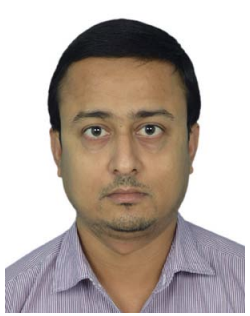}}]{Nur Al Hasan Haldar}
received the Ph.D. degree in computer science from the University of Western Australia (UWA). He is currently a Lecturer with the School of Electrical Engineering, Computing, and Mathematical Sciences (EECMS), Curtin University, Australia. His research interests include graph data analytics and optimization, query processing, social network analytics, cybersecurity, and machine learning.
\end{IEEEbiography}
\vspace{-10mm}
\begin{IEEEbiography}[{\includegraphics[width=1in,height=1.25in,clip,keepaspectratio]{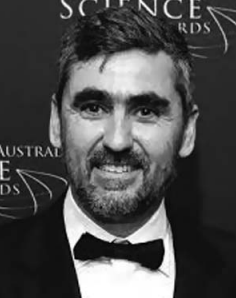}}]{Ajmal Mian}
(Senior Member, IEEE) is currently a Professor of Computer Science with the University of Western Australia. His research interests include 3D computer vision, machine learning, human action recognition, and video description. He is a Fellow of the International Association of Pattern Recognition. He has received three prestigious fellowships from the Australian Research Council (ARC) and several awards, including the West Australian Early Career Scientist of the Year Award, the Aspire Professional Development Award, the Vice-Chancellor's Mid-Career Research Award, the Outstanding Young Investigator Award, the IAPR Best Scientific Paper Award, the EH Thompson Award, and the Excellence in Research Supervision Award. He has received several major research grants from the ARC, the National Health and Medical Research Council of Australia, and the U.S. Department of Defense. He serves as a Senior Editor for IEEE Transactions on Neural Networks and Learning Systems, and as an Associate Editor for IEEE Transactions on Image Processing and the Pattern Recognition journal.
\end{IEEEbiography}

\end{document}